# Tailored RF Pulse for Magnetization Inversion at Ultrahigh Field

Aaron C. Hurley,[i,2] Ali Al-Radaideh,[i] Li Bai,[2] Uwe Aickelin,[2] Ron Coxon,[i] Paul Glover,[i] and Penny A. Gowland[i*]

The radiofrequency (RF) transmit field is severely inhomogeneous at ultrahigh field due to both RF penetration and RF coil design issues. This particularly impairs image quality for se- quences that use inversion pulses such as magnetization pre- pared rapid acquisition gradient echo and limits the use of quantitative arterial spin labeling sequences such as flow-at- tenuated inversion recovery. Here we have used a search algo- rithm to produce inversion pulses tailored to take into account the heterogeneity of the RF transmit field at 7 T. This created a slice selective inversion pulse that worked well (good slice profile and uniform inversion) over the range of RF amplitudes typically obtained in the head at 7 T while still maintaining an experimentally achievable pulse length and pulse amplitude in the brain at 7 T. The pulses used were based on the frequency offset correction inversion technique, as well as time dilation of functions, but the RF amplitude, frequency sweep, and gradient functions were all generated using a genetic algorithm with an evaluation function that took into account both the desired inversion profile and the transmit field inhomogeneity.  Magn Reson Med 63:51–58, 2010.

The radiofrequency (RF) transmit field is severely inhomogeneous at ultrahigh field due to both RF penetration and RF coil design issues. This is a particular problem for techniques that use inversion pulses such as magnetization prepared rapid acquisition gradient echo (MPRAGE) (i) and arterial spin labeling sequences such as flow attenuated inversion recovery (2). A number of approaches to solving this problem are being investigated, generally requiring additional hardware. Here we have taken a simpler approach, using a search algorithm to produce inversion pulses tailored to take into account the heterogeneity of the RF transmit field at 7 T. The goal was to create a slice selective inversion pulse that worked well (good slice profile and low sensitivity to RF inhomogeneity) over the range of RF amplitudes typically obtained in vivo while still maintaining an experimentally achievable pulse length and pulse amplitude in the brain at 7 T.

The standard inversion pulse used in MRI is the hyperbolic secant inversion pulse (3), which has previously been subject to reshaping (frequency offset corrected inversion [FOCI] pulses) to provide improved slice profile over a wide range of pulse powers (4) and to time resampling/dilation (variable rate selective excitation) to reduce the pulse power or improve inversion at low RF amplitudes (5,6). In this work, the RF amplitude, frequency sweep, and gradient functions were optimized using a genetic algorithm (7) with an evaluation function that took into account both the desired inversion slice profile and sensitivity to transmit field inhomogeneity, within the constraints of maximum RF field amplitude ($B_i$) and fixed pulse length. We used this to optimize two similar pulses. The first pulse is the C-shape FOCI (C-FOCI) pulse, which is defined by three variables ( , , and $A_{max}$), with no time resampling. The second pulse we call the time resampled FOCI (TR-FOCI) pulse, which uses a time resampling function and a more general reshaping functions and is defined by ii variables.

A genetic algorithm was used to search a large space of possible reshaping and resampling functions to find tailored solutions for ultrahigh field, with an evaluation function (describing the features being optimized) related to slice profile and sensitivity to $B_i$ nonuniformity. Similar work in the past (8) optimized the amplitude at a set of evenly spaced points within the wave form, the rest of the points being filled in by cubic spline interpolation before evaluation. Our system differs in that it uses the FOCI approach, so that the amplitude and frequency modulation functions are not optimized directly, but instead their reshaping and resampling functions are optimized. The RF pulses are initially defined by a pair of orthogonal amplitude and frequency modulation envelopes, e.g., (sech,tanh). These functions are defined over the range of normalized time ($-i < t < i$), and time resampling changes the sampling of the functions from uniform intervals to variable intervals. Our method maintains symmetry of "time" about zero and the starting and end points are held fixed.

The aim of this work was to produce inversion pulses for different applications at 7 T. Pulses with various slice thicknesses were produced that would be suitable for three-dimensional imaging (with MPRAGE), and arterial spin labeling. Short pulses were also optimized. This paper outlines the genetic algorithm used to optimize the pulses, and in particular the evaluation function used. It then presents the optimized pulses and their simulated and experimental behavior.

## THEORY

C-FOCI pulses (4) can be generated from the three parameters contained in the vector [$A_{max}$   ] chosen via a search

[1] Sir Peter Mansfield Magnetic Resonance Centre, University of Nottingham.
[2] Department of Computer Science, University of Nottingham.

Grant sponsor: EU Fp6 Marie Curie Action Programme; Grant number: MEST-CT-2005-021170.

*Correspondence to: P. A. Gowland, XX, Sir Peter Mansfield Magnetic Resonance Centre, University of Nottingham, NG7 2RD, UK. E-mail: penny.gowland@nottingham.ac.uk

Received 4 April 2009; revised 23 June 2009; accepted 17 July 2009.
DOI 10.1002/mrm.22167

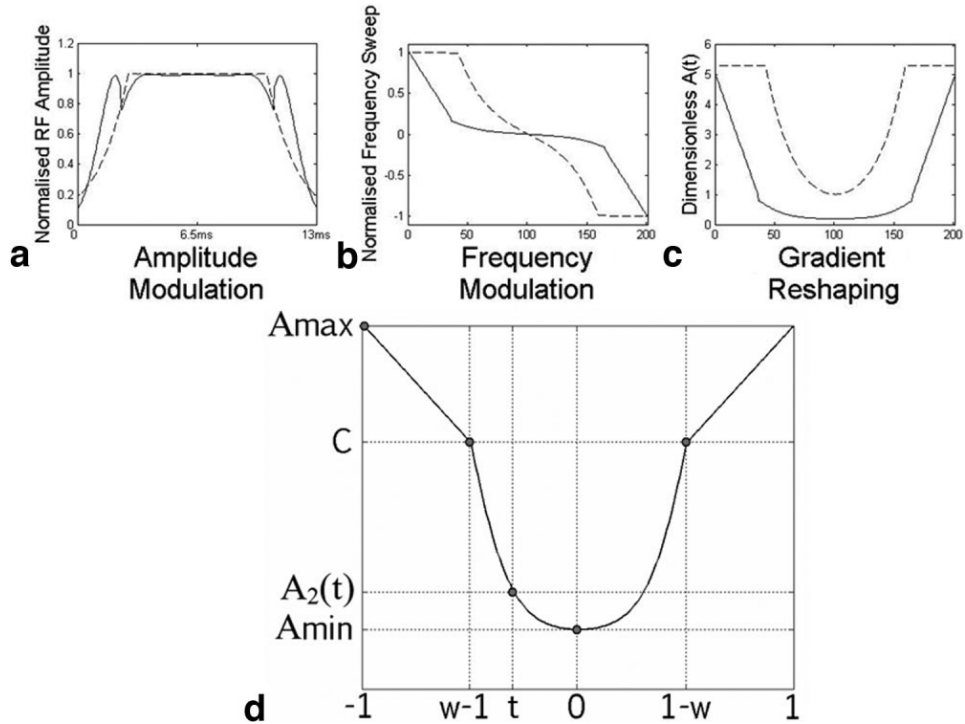

FIG. 1. A C-FOCI (dotted line) and a TR-FOCI pulse (solid line). **a:** RF amplitude, **(b)** RF frequency, and **(c)** gradient amplitude modulation functions. **d:** Schematic diagram illustrating the meanings of the parameters used to define the gradient reshaping function.

algorithm. TR-FOCI pulses are based on FOCI pulses with time resampling (6) and can be generated from the 11 parameters contained in the vector $[A_{max}, w, r_1, r_2, r_3, r_4, r_5, \ldots, T_i, T_2]$. The first seven terms define the shaping function, and are parameters inherited from FOCI pulses, and $T_i$ and $T_2$ define a time resampling function. Typical pulse profiles (RF amplitude, RF frequency modulation, and gradient amplitude) are shown in Fig. 1a–c. The pulses are optimized using a genetic algorithm, and to do this we need to define an evaluation function, to be described later, which we have termed the weighted inversion profile accuracy (WIPA) function.

### Simulation Program

In order to be able to evaluate and optimize the performance of a pulse, its effect on the magnetization must be simulated. For this we used a program that simulated rotations in the rotating frame due to the $B_1$ amplitude and off resonance, but not relaxation. The pulses were sampled at 200 points corresponding to the experimental sampling intervals. The profile was estimated at 1000 different positions over a width $M$ mm, with the desired inversion slab being of thickness of $M/2$ mm at the center of the profile. The magnetization was assumed to be at equilibrium before the pulse.

### Evaluation Function

The evaluation function characterizes both the slice profile of the pulse and its sensitivity to RF amplitude variations. We define an inversion profile accuracy (IPA) function:

$$IPA = \frac{1000}{\sum_{i=-500}^{500} (I(i) - V(i))^2} \quad [1]$$

where $i$ is the index corresponding to distance (in sample points) from the center of slice profile, $I$ is the ideal profile which is set to $+1$ for $i - 500 < 251$ and $-1$ elsewhere, and $V$ is the actual inversion profile as determined from the simulation. Figure 2a shows the ideal inversion profile and a representative actual inversion profile.

Figure 2b demonstrates the variation in IPA with $B_1$ amplitude for two different pulse shapes. It can be seen that although the pulse that generated the dashed line gives a higher peak value of IPA, the pulse that generated the solid line performs more consistently over a range of different $B_1$ amplitudes. However, it would be impracticable to evaluate every candidate pulse at many RF amplitudes, as shown in Fig. 2b, so instead the IPA was evaluated at three values, a lower limit $L_1 = 3$ μT, an upper limit $L_3 = 7$ μT, and the midpoint $L_2 = \frac{L_1 + L_3}{2}$ (see Fig. 1e), and then a WIPA was calculated:

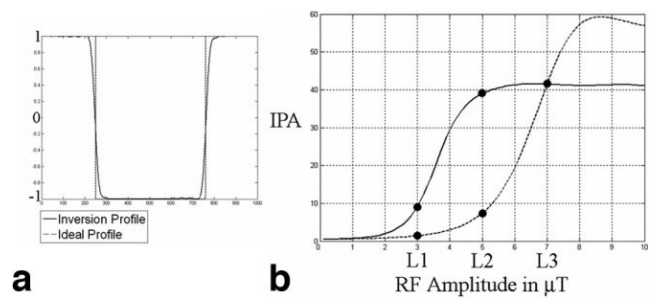

FIG. 2. **a:** Comparison of the ideal (solid) with an actual (dashed) inversion slice profile. **b:** Comparison of the IPA functions for two pulses.

$$WIPA(L_1, L_3) = IPA(L_1)\frac{L_3}{L_1 + L_2 + L_3} + IPA(L_2)\frac{L_2}{L_1 + L_2 + L_3}$$
$$+ IPA(L_3)\frac{L_1}{L_1 + L_2 + L_3} \quad [2]$$

The weighting emphasizes performance at the lower limit since pulses generally perform better at higher RF amplitudes. Initial experiments showed that if the WIPAs were calculated at only two points, this tended to produce pulses with IPA peaking at those points, whereas the inclusion of the third point, $L_2$, tended to produce smoother curves (rather than three peaks). The RF amplitude values 3, 5, and 7 uT were chosen as representative of very low RF amplitudes in vivo.

It will be explained further below that for pulses that achieved an acceptable value of WIPA, the full IPA curves (Fig. 2b) were calculated, and the area under the full IPA curves was found (integrated IPA). The pulse giving the maximum value of integrated IPA was selected as optimum.

### Shaping Functions

The flip angle of an adiabatic pulse is determined by the frequency modulation, rather than amplitude modulation of the pulse, provided that the pulse meets the adiabatic condition. FOCI pulses are based on the hyperbolic secant pulse but include additional gradient modulation with parallel reshaping of the amplitude modulation and frequency modulation envelopes to increase the periods of the pulse for which the adiabatic condition holds for a given resonance offset (4). They can be defined by the following equations:

$$B_1(t) = A(t)\text{sech}(\beta t)$$
$$\Delta\omega(t) = -A(t)\mu\beta\tanh(\beta t) \quad [3]$$
$$G(t) = A(t)G_s$$

where $B_1(t)$ is the final amplitude modulation function, $zw(t)$ is the frequency modulation function, and $G(t)$ is the gradient modulation function. The sech and tanh functions are the underlying modulation functions of the hyperbolic secant pulse, $G_s$, is the time-constant slice selection gradient, and $A(t)$ reshapes all the modulation functions. $A(t)$ is continuous and symmetric and monotonically increases with $t$.

The C-FOCI pulse modulation function (A) consists of two constant segments joined by a curved segment, which is defined by the cosh function, with the length of each segment being a function of the max value of the gradient shape (dotted line in Fig. 1c). The TR-FOCI pulse also uses a modulation function defined by three segments, but the first ($A_1$) is linear, the curved section ($A_2$) is defined by a polynomial of even terms, and the third segment is a mirror image of $A_1$ (solid line in Fig. 1c). For the TR-FOCI pulse, the shaping function is defined by the first seven terms of the vector $[A_{max}, w, r_1, r_2, r_3, r_4, r_5, u, t, T_i, T_2]$ (Fig. 1d). For negative $t$, the linear segment is described by three parameters: the length $w$, the starting height of the shaping function $A_{max}$, and a parameter $r_1$, which determines the slope, which is given by $-\frac{r_1 A_{max}}{w}$. Therefore, the value of the linear segment at the end point ($t = w - 1$) is $c = A_{max}(1 - r_1)$, and the linear segment is described by

$$A_1(t) = A_{max}\left(1 - \frac{r_1(t+1)}{w}\right) \quad [4]$$

The curved segment is described by a polynomial with even powers and positive coefficients of order 8 for the sake of efficiency of the search:

$$A_2(t) = b(1)t^2 + b(2)t^4 + b(3)t^6 + b(4)t^8 + A_{min} \quad [5]$$

The inputs $r_2, r_3, r_4, r_5$ determine the coefficients of the polynomial so that:

$$A_{min} = r_2 c \quad (c >= A_{min} = 0) \quad [6a]$$
$$b(1) = r_3\frac{c - A_{min}}{(w-1)^2} \quad [6b]$$
$$b(2) = r_4(1 - r_3)\frac{c - A_{min}}{(w-1)^4} \quad [6c]$$
$$b(3) = r_5(1 - r_4)(1 - r_3)\frac{c - A_{min}}{(w-1)^6} \quad [6d]$$
$$b(4) = (1 - r_5)(1 - r_4)(1 - r_3)\frac{c - A_{min}}{(w-1)^8} \quad [6e]$$

ensuring that $A_2(w - 1) = c$ and so continuity of A, which is essential for a practical gradient waveform (one which does not exceed the maximum allowed slew rate).

### Time Resampling

Given a resampling function $T(t)$, the equations describing the new pulse become:

$$B_1(t) = A(t)\text{sech}(\beta T(t))$$
$$\Delta\omega(t) = -A(t)\mu\beta\tanh(\beta T(t)) \quad [7]$$
$$G(t) = A(t)G_s$$

$T$ is defined to be monotonically increasing from $-1$ to 1. We limited our search to polynomials of odd powers of order 5:

$$T(t) = \frac{\tau_1 t^5 + \tau_2 t^3 + t}{\tau_1 + \tau_2 + 1} \quad [8]$$

The coefficient for $t^1$ is redundant since the function is scaled to remain in the range $[-1,1]$. The parameters were constrained in the following ranges: $1 < A_{max} < 30$, $0.5 < u < 10$, $1 < t < 10$, $[0 < r_i < 1 \ (i = 1,2,3,4)]$, $[0 < T_i < 5, i =$

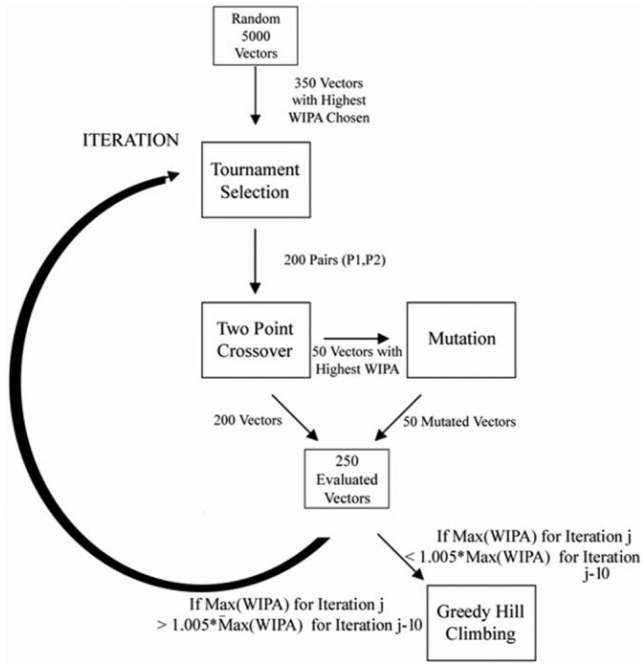

FIG. 3. Diagram illustrating the genetic algorithm as used for optimizing the TR-FOCI pulse.

1,2]. An example of one vector that gave good early results is [4.94 0.36 0.84 0.24 0.31 0.80 3.99 6.07 6.10 0.28 1.35]. The limitations are set to narrow the search space and were informed by exploratory data. There are further hardware limits that must also be taken into account, and these are discussed below.

The following description of the genetic algorithm is for TR-FOCI pulse, but the C-FOCI pulse was optimized by a similar process.

### Optimization With a Genetic Algorithm

We generated an initial population of vectors by taking 350 vectors with the highest WIPA from 5000 randomly generated vectors (within allowed ranges for each parameter). There were then two stages to the genetic algorithm (Fig. 3).

### Iteration

One iteration uses four independent tournament selections, with the 200 (50 X 4) output pairs subjected to two-point crossovers, whose outputs are passed to the next iteration. Mutations of the 50 two-point crossover vectors with the highest WIPA are also passed on to the next iteration.

A number of iterations are run, with the output from the previous iteration being the input for the current iteration. Once there is no discernable improvement from the previous 10 iterations (<0.5% increase in maximum WIPA), the algorithm passes onto the next phase, greedy hill climbing. We first describe the three processes involved to complete one iteration:

*Two-point crossover.* This is the process whereby the two vectors $P_1$ and $P_2$ are combined. Given two crossover points $m_1$ and $m_2$, a new vector $Q$ is created according to $Q = [P_1(1: m_1 - 1)\ P_2(m_1: m_2 - 1)\ P_1(m_2:\text{end})]$. Often in point crossover, the points are chosen randomly. However, in this case the vector $[A_{max}\ w\ r_1\ r_2\ r_3\ r_4\ r_5\ u\ t\ T_1\ T_2]$ naturally falls into four subsections describing distinct aspects of the pulse, and within which the elements are highly connected: elements 1 to 3 determine the line segment of the reshaping function, elements 4 to 7 determine the other aspects of the reshaping function, elements 8 and 9 relate to the frequency sweep, and elements 10 and 11 are time resampling parameters. Therefore it is efficient to keep these groups of elements together and perform two-point crossover at ($m_1 = 4$ and $m_2 = 8$), ($m_1 = 4$ and $m_2 = 10$), ($m_1 = 8$ and $m_2 = 10$), and ($m_1 = 3$ and $m_2 = 8$ since element 3 determines the start of the curve described by elements 4-7). Two hundred (4 X 50) new vectors are generated in this manner. The resulting 50 vectors with the highest WIPA values are then mutated.

*Mutation.* Given a vector $W_i$, a new vector is produced, $U_i$, such that:

$$U_i = W_i(1 + \text{Ra}(1 - (1/b))^a) \qquad [9]$$

where $a$ ($0 < a < 1$) and $b$ ($1 < b < \infty$) are variables that control the size of the mutation and R is a random variable ($-1 < R < 1$). A series of tests of a series of discrete values of $a$ and $b$ found that $a = 0.05$ and $b = 50$ resulted in the most dynamic move upwards in WIPA.

*Tournament selection.* This process takes 250 vectors (350 for the initial population) that have all been evaluated and divides them randomly into 50 groups of five vectors (seven for the initial population). The "winner" of the "tournament" is the vector with the highest evaluation within each group and is chosen as the "first parent" $P_1$. The second parent, $P_2$, is chosen randomly from the remaining four or six losers. The 50 ($P_1,P_2$) pairs are then used in two-point crossover.

### Greedy Hill Climbing

This procedure was performed on the 10 best vectors from the final iteration independently. For a given vector $[A_{max}\ w\ r_1\ r_2\ r_3\ r_4\ r_5\ u\ t\ T_1\ T_2]$, one parameter at a time was changed within its allowed range, leaving the rest of the vector constant, and the value of that parameter that gave the highest WIPA was chosen. This is then repeated for each parameter of the vector in turn. The values tested were chosen at random intervals in an allowed range (the number of values tested was a function of the size of the range). This search of each of the vector's parameters in turn was repeated until no significant improvement occurred (<0.5% increase in WIPA).

The final vector is chosen from the 10 vectors that went through greedy hill climbing by numerically integrating the IPA curves (Fig. 2b) from 0 to 10 uT at 0.1-uT intervals. The vector yielding the highest integral is chosen to generate the TR-FOCI pulse.

### MATERIALS AND METHODS

The genetic algorithm was used to design TR-FOCI and C-FOCI pulses to invert 1-, 5-, and 50-mm slice thickness, for a maximum gradient amplitude of 33 mT/m and a length T = 13 ms, over a $B_1$ range of 3-15 uT, corresponding to the range observed in vivo. A shorter TR-FOCI pulse was also designed for practical non–slice selective (200 mm) inversion for a pulse length of 5 ms, but for this pulse the evaluation function was modified to give less weight to the sharpness of the slice profile. A hyperbolic secant (HSC) pulse (T = 13 ms, u = 6.25, t = 4.5) was also simulated to match the pulse that was standard on the scanner.

Table 1
Vectors of Parameters Defining the Optimum C-FOCI and TR-FOCI Pulses for a Range of Different Slice Thicknesses and Pulse Lengths

| Slice thickness (mm) | Pulse length (ms) | C-FOCI | TR-FOCI |
|---|---|---|---|
| 1 | 13 | (1.43 5.46 3.64) | (2.88 0.52 0.46 0.45 0.68 1.00 0.29 2.33 4.27 0.10 0.40) |
| 5 | 13 | (1.67 5.07 17.14) | (3.99 0.28 0.67 0.11 0.22 0.09 0.96 5.83 5.82 0.14 1.21) |
| 50 | 13 | (19.8 5.41 1.45) | (5.98 0.32 0.73 0.18 0.33 0.81 0.04 4.9 4.90 0.00 0.83) |
| 200 | 5 | (21.87 4.25 2.15) | (3.32 0.30 0.64 0.27 0.59 0.00 1.00 7.71 3.90 0.25 0.40) |

The maximum gradient strength used in the pulse was set to an initial value of $G'_{max}$ = 1.15 $A_{max}ut$/SL (where SL = slice thickness), and then the required value of $G_{max}$ was found by multiplying $G'_{max}$ by the ratio of simulated slice thickness (width of profile at zero crossing for initial value of $G'_{max}$) to the desired slice thickness. If this re- sulted in $G_{max}$ > 33 mT/m (the maximum gradient avail- able on the system), then $G_{max}$ was limited to be 33 mT/m, thus punishing pulses unable to achieve the desired slice thickness for practical gradient strength restrictions. There are also hardware limits on the maximum frequency sweep of 50 kHz, which has the effect of limiting the product of *Amax*, u, and t to 2050.

Simulations were used to compare the inversion profiles of three different pulses (TR-FOCI, C-FOCI and HSC) over the range of $B_1$ amplitudes 4-13 uT. To test the pulses experimentally, they were used to invert a slice perpen- dicular to the imaging plane of an EPI scan, at the isocenter of the magnet/gradient system. The gradient waveform amplitude was adjusted to set the slice thickness to 30 mm (in the case of pulses for 5-mm inversion) to provide ade- quate resolution across the profile. Experiments were per- formed on a 7-T Philips Achieva scanner with a 16-chan- nel Nova Medical brain receive coil, and using a saline-filled, spherical phantom. Data were acquired at a variety of inversion times and a long pulse repetition time. The resulting images could be fitted to S(inversion time) = So(1 − Ke$^{-inversion time/T1}$) on a pixel-by-pixel basis for So, $T_1$ and K, so that the resulting maps of K gave the inversion efficiency across the slice profile. This was carried out not only for the standard $B_1$ amplitude used for the C-FOCI pulses but also for a range of lower amplitudes to simulate the effects of RF heterogeneity.

We also compared the use of the optimized pulses to provide a slab inversion in whole-head brain MPRAGE scans (pulse repetition time = 15 ms, shot-to-shot intervals 3000 ms, turbo field echo (TFE) factor = 148, echo time 5.9, inversion time = 1050 ms, flip angle = 8, resolution 0.6 mm isotropic, matrix size 320, radial *k*-space sampling) in several subjects.

Further simulations were carried out to investigate the effect of gradient amplifier nonidealities or eddy currents on pulse performance by convolving the gradient wave- forms with an exponential function with a decay constant of 5 sec$^{-1}$. The performance of the pulses far from reso- nance was also investigated.

## RESULTS

Table 1 shows the parameters for the optimum C-FOCI and TR-FOCI pulses inverting various slice thicknesses and pulse lengths. Figure 4 shows the simulated slice profiles for the different pulses at a range of RF amplitudes (over the range encountered in vivo at 7 T). For the 1-mm slice thickness, 13-ms pulse, it can be seen that TR-FOCI performs well over a broad range of RF amplitude, whereas the C-FOCI performs well at medium to low RF amplitudes but performs poorly at high RF amplitude (the C-FOCI could be optimized for high rf amplitude but only at the expense of its low RF amplitude performance). For such a thin slice, the limits on gradient strength mean that the bandwidth of the pulse must be low, which limits the product of $A_{max}$, u and t, but the pulse performance (in terms of slice profile) is reduced if $A_{max}$ is limited. The profiles for the hyperbolic secant are shown for comparison and are the same for all slice thicknesses since the pulse defined on the scanner is a relatively low bandwidth pulse and so does not exceed the available gradient even for a 1-mm slice. Figure 4 also shows that for the 5-mm slice, 13-ms pulse, the C-FOCI pulse performs better than the HSC pulse for $B_1 < 10$ uT, although the profile becomes increasingly distorted as the amplitude falls very low. In contrast, the TR-FOCI pulse retains a good slice profile until very low RF amplitudes. At high $B_1$ amplitudes >10 mT, all pulses perform well, with the C-FOCI pulse providing the optimum slice profile. Figure 5 shows that this TR-FOCI pulse also gave a good slice profile experimentally, with low sensitivity to reduced $B_1$ amplitude. However the experimental inversion profiles for all pulses are somewhat more distorted than predicted by the simulations, particularly for the C-FOCI pulse. Figure 4 also shows that the 50-mm slice, 13-ms pulse shows similar behavior to that for the 5-mm-slice pulse: the C-FOCI profile started to degrade at 4 uT, whereas the TR-FOCI continued to invert with a very good profile at this RF amplitude. Figure 6 demonstrates the use of these pulses in vivo. The HSC pulse performed very poorly, as expected in regions on the brain where RF inhomogeneities are strong (e.g., at the temporal lobes in Fig. 6). The C-FOCI pulse led to loss of gray matter/white matter (GM/WM) contrast in the cerebellum (see Fig. 6b), whereas the TR-FOCI pulse retained GM/WM contrast in the cerebellum. Figure 7 shows the simulated profiles for the 200-mm, 5-ms pulses at different RF amplitudes, demonstrating that a stable inversion could be achieved for very short pulse using an optimized TR-FOCI pulse.

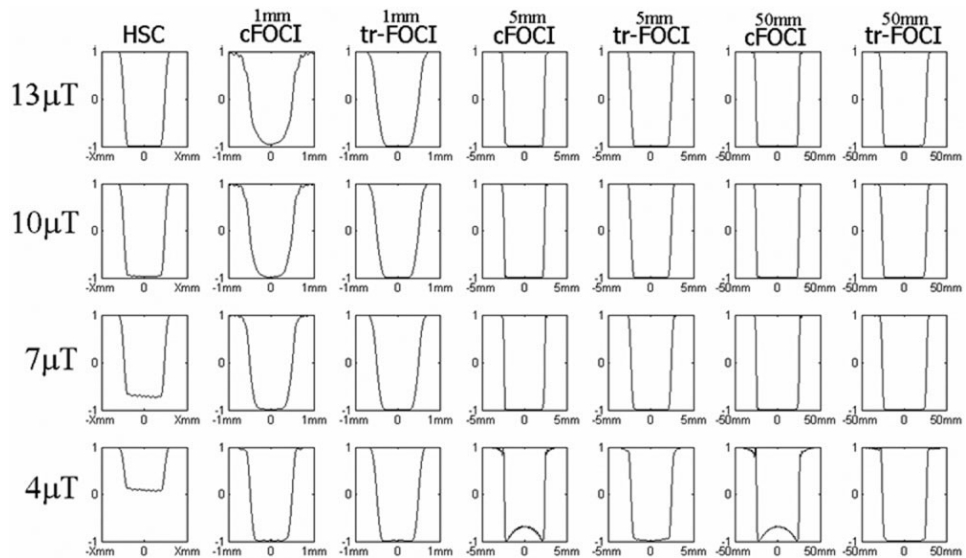

FIG. 4. Simulated inversion pro- files for HSC, C-FOCI, and TR- FOCI pulses at various level of RF amplitude and for differing slice thicknesses.

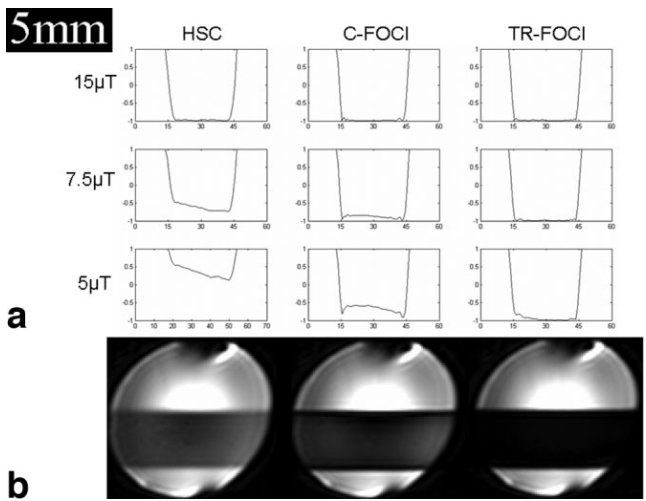

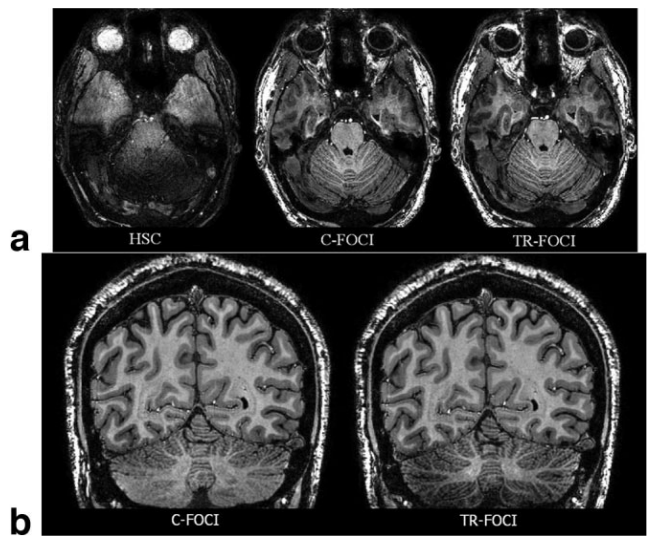

FIG. 5. **a:** Experimental inversion profiles for HSC, C-FOCI, and TR-FOCI pulse shapes for nominal RF amplitudes 15, 7.5, and 5 uT. **b:** Scans at 7 T of phantom near the null point for inversion at a nominal RF amplitude of 5 uT for HSC, C-FOCI, and TR-FOCI pulses.

FIG. 6. **a:** Transverse MPRAGE images using HSC, C-FOCI, and TR-FOCI for inversion. **b:** Coronal MPRAGE images for C-FOCI and TR-FOCI pulses, showing improved contrast at base of brain for TR- FOCI.

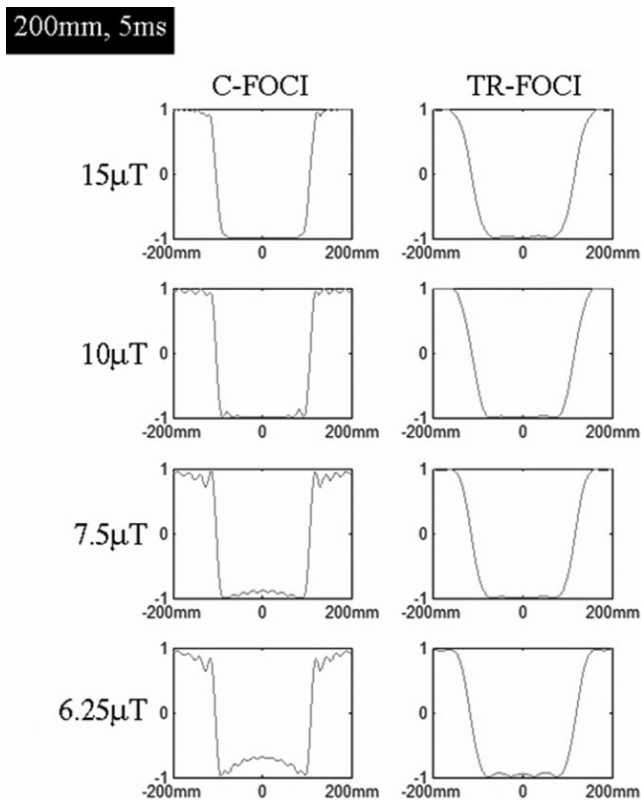

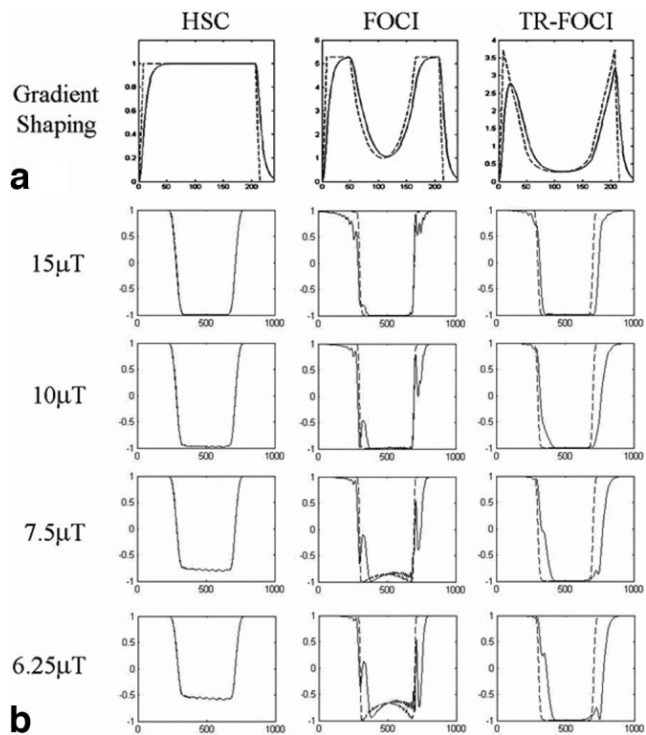

FIG. 7. Simulated inversion profiles for HSC, C-FOCI, and TR-FOCI pulse shapes for nominal RF amplitudes 15, 7.5, and 5 uT.

FIG. 8. **a:** The original gradient shapes (–) and a distorted equivalent. **b:** Inversion profiles at different levels of RF amplitude, without (–) and with gradient distortion.

Convolving the gradient waveform with an exponential function with a decay constant of 5 sec$^{-1}$ designed to simulate the effects of eddy currents or nonideal amplifiers resulted in errors in the inverted slice profile for both a C-FOCI and TR-FOCI but had little effect on the profile of the HSC pulse (see Fig. 8). The effect was worse for the C-FOCI than the TR-FOCI pulse. Further simulations showed that errors in the gradient waveform during the periods when the RF amplitude was high had a more significant effect on the slice profile than errors at the start of the pulse when the RF amplitude is changing slowly and the adiabatic condition is easily met.

Figure 9 shows that off resonance, at very high power the C-FOCI pulse shows significant side bands that are less evident with the TR-FOCI pulse.

## DISCUSSION

A genetic algorithm has been used to design a range of slice selective inversion pulses that can perform well across the range of $B_1$ amplitudes currently achievable in the human brain at 7 T with commonly available hardware, producing pulses suitable for three-dimensional imaging and arterial spin labeling. Both standard C-FOCI and TR-FOCI pulses were investigated, and it was found that, for the evaluation function used here, the C-FOCI provided the sharpest (most square) slice profile at high RF power, but the performance of the TR-FOCI pulse was most robust to variations in $B_1$ amplitude. The pulses designed here could also be adapted for use at larger slice thicknesses by scaling the slice select gradient, although with increased sensitivity to field inhomogeneity. This builds on the previous work of Yongbi et al. (9) and Ordidge et al. (4), who proposed the use of a high value for $A_{max}$ (10-10.5) for the C-FOCI pulse to give a sharper profile, with reasonable robustness to variations in $B_1$ amplitude, for larger slice thickness.

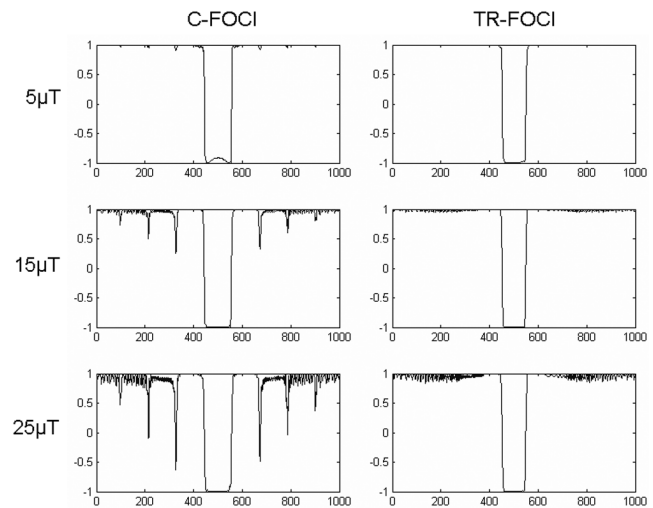

FIG. 9. Inversion profiles for C-FOCI and TR-FOCI pulses showing distortion beyond the slice region for very high RF amplitudes, particularly for C-FOCI.

The TR-FOCI variables can also be used to describe the C-FOCI pulse by approximating the cosh function used to describe the shaping function by a Maclaurin expansion, calculating the width of the linear segment as a function of $A_{max}$ and setting its slope factor ($r_1$) to zero and the time resampling coefficients to zero. In other words, the approximated C-FOCI pulses can be considered to be a subset of the TR-FOCI pulses and so were contained within the search space of the genetic algorithm optimization of the TR-FOCI pulses. Therefore, since the result of the optimization of the TR-FOCI pulse was not a C-FOCI pulse, this implies that the C-FOCI pulses are not optimal for the evaluation function chosen here.

Clearly, the optimization (and in particular the evaluation function) should be tailored for the specific application (e.g., slab inversion in MPRAGE sequences compared to thin slice inversions for arterial spin labeling), hardware and sequence limits (maximum gradient amplitude and pulse length), and specific experimental conditions (range of $B_1$ and amplitude of static polarizing field [$B_0$] inhomogeneities expected). The tradeoff between edge sharpness and robustness to variations in $B_1$ amplitude can be altered by adjusting the weighting of the edge and center of the profile in the evaluation function.

Simulations showed that reducing the length (T) of TR-FOCI, C-FOCI or HSC pulses made the slice profile less sharp and the pulse more sensitive to RF inhomogeneity, although this effect was less pronounced for TR-FOCI, and in fact it was possible to produce a very robust TR-FOCI pulse that could invert for a low integrated RF power.

The simulations took no account of $B_0$ inhomogeneity, which is particularly problematic at high field and will change the performance of the pulse lead to distortion of slice profiles in arterial spin labeling (10) and also in saturation bands. Given the subject-specific nature of the $B_0$ field inhomogeneities and the spatial variation in the $B_0$ field gradients, it is not feasible to include the effects of these field gradients in the optimization. However, they can be minimized by designing low-bandwidth pulses to be used with higher gradient amplitudes. These simula- tions took also no account of the effect of relaxation during the pulse, which would alter the performance particularly for the longer pulses.

The simulations did not limit the gradient slew rate, nor take account of nonidealities in the gradient waveform due to eddy currents or unexpected performance of the gradi- ent amplifiers, but due to the nature of the gradient wave- forms, the slew rates for all pulses mentioned are well below the specification of the gradient system. Further simulations showed that nonidealities of the gradient waveforms had little effect on the HSC pulse (as expected) but resulted in errors in the inverted slice profile for both a C-FOCI and TR-FOCI, although this effect was worse for the C-FOCI than the TR-FOCI pulse (see Fig. 8). Further simulations showed that errors in the gradient waveform during the periods when the RF amplitude was high have more effect on the slice profile than errors at the start of the pulse when the RF amplitude is changing slowly and the adiabatic condition is easily met. Furthermore, although simulations suggested that the C-FOCI pulse should give a better slice profile than TR-FOCI at high RF amplitudes, experimentally it was found that the C-FOCI pulse gave distorted slice profiles when the slices were inverted away from isocenter (results not presented), and we suggest that this effect is due to the effect of eddy currents or other errors in the gradient waveform. Since the TR-FOCI pulse gradient reshaping function is smoother than that for the C-FOCI pulse, it is expected to be less sensitive to any nonidealities in the gradient waveforms. If the gradient performance were fully characterized, then this could be used as an additional parameter in the search algorithm. This may require either a weighted measure of importance for gradient distortion in the evaluation function or the setting of a multiobjective optimization problem (11).

It can be seen that in general the pulses perform better at higher $B_1$ amplitude. However, the maximum $B_1$ available is generally limited by the RF coil, and even if this were overcome, for instance, by using local transmit coils, ulti- mately the performance of RF pulses breaks down at higher $B_1$ amplitudes. This can be seen in Fig. 4 for the 1-mm C-FOCI pulse, which starts to exhibit side bands at high $B_1$, and also in Fig. 9. Such side bands could cause image artifacts in some sequences, and where relevant, future optimizations should use evaluation functions that take into account the performance of the pulse well be- yond the slice thickness.

## CONCLUSION

We have designed inversion pulses capable of performing well across a range of $B_1$ amplitudes and in particular at low RF amplitude. The search algorithm addressed the issue of inhomogeneity in the $B_1$ field, but not inhomoge- neity the $B_0$ field, or problems that exist in the reproduc- tion of the desired gradient waveforms. The robustness of these pulses to RF inhomogeneities has made routine use of MPRAGE in the temporal lobes possible on our 7-T scanner.

## ACKNOWLEDGMENTS

This work was funded by the Medical Research Council, UK. The University of Nottingham 7-T scanner was origi- nally funded by Wellcome Trust, the Office of Science and Technology and the Higher Education Funding Council for England.